\documentclass[3p,times,twocolumn]{elsarticle}

\usepackage{lineno}
\usepackage{xcolor}   
\usepackage{ulem}
\usepackage{graphicx}

\def\um{\ifmmode {\mathrm{\mu m}}\else
                  \textrm{$\mu$m }\fi}%
\def\GeV{\ifmmode {\mathrm{\ Ge\kern -0.1em V}}\else
                   \textrm{Ge\kern -0.1em V}\fi}%
\def\MeV{\ifmmode {\mathrm{\ Me\kern -0.1em V}}\else
                   \textrm{Me\kern -0.1em V}\fi}%
\def\keV{\ifmmode {\mathrm{\ keg\kern -0.1em V}}\else
                   \textrm{ke\kern -0.1em V}\fi}%
\def\eV{\ifmmode  {\mathrm{\ e\kern -0.1em V}}\else
                   \textrm{e\kern -0.1em V}\fi}%

\def\uW{\ifmmode  {\mathrm{\mu  W}}\else
                   \textrm{$\mu$W}\fi}%

\begin{document}
\begin{frontmatter}

\title {
Prototyping of a 25 Gbps optical transmitter for applications in 
high-energy physics experiments
}

\author[apac]{C.-P. Chao}
\author[apac]{S.-W. Chen}
\author[smu]{D. Gong}
\author[ipas]{S. Hou}
\author[smu]{X. Huang} 
\author[apac]{C.-Y. Li}
\author[smu]{C.~Liu}
\author[smu]{T.~Liu}
\author[nju]{M.~Qi}
\author[smu]{J.~Ye}
\author[smu]{X.~Zhao}
\author[nju]{L.~Zhang}
\author[smu]{W.~Zhou}

\address[apac]{  APAC Opto Electronics Inc., Hsinchu, Taiwan 303 }
\address[smu]{   Southern Methodist University, Dallas, TX 75275  }
\address[ipas]{  Academia Sinica, Taipei, Taiwan 115 }
\address[nju]{   Nanjing University, Nanjing 210093, China }

\begin{abstract}

Development of optical links with 850 nm multi-mode vertical-cavity 
surface-emitting lasers (VCSELs) has advanced to 25 Gbps in speed.
For applications in high-energy experiments,  the transceivers are 
required  to be tolerant  
in radiation and particle fields. 
We report on prototyping of a miniature transmitter named MTx+, which is developed 
for high speed transmission with the dual-channel laser driver LOCld65 and 
850 nm VCSELs packaged in TOSA format. 
The LOCld65 is fabricated 
  in  the TSMC 65 nm process and is packaged 
in the QFN-40 for assembly. 
The MTx+ modules and test kits were first made with PCB and components qualified
for 10 Gbps applications, and   were 
tested for achieving 14 Gbps.
The data transfer rate of the MTx+ module is investigated further for the speed of 
up to 25 Gbps.
The LOCld65 is examined with post-layout simulation and the module design upgraded 
with components including the TOSA qualified for 25~Gbps applications.
The PCB material is replaced by the Panasonic MEGTRON6.  
The revised MTx+ is tested at 25 Gbps and the 
eye-diagram shows a mask margin of 22~\%.

\end{abstract}
\end{frontmatter}

\section{Introduction}

Optical links with 850 nm multi-mode fiber and Vertical-Cavity 
Surface-Emitting Laser (VCSEL) provide the advantage of high speed 
data transmission over a distance of a few hundreds meters.
When applied in high energy experiments, the opto-electronics are 
required   to be radiation hard.  
The VCSEL structure is commonly made of a thin implantation of about 10 nm
on a GaAs wafer.
The radiation effects to VCSELs have been studied for
tolerance to ionizing dose and particles fluence
\cite{VCSEL_radhard}. 
The laser driver ASIC requires customization for functionalities
specified by the experiments and the circuit design to withstand 
radiation.

In the following we report on the MTx+, a dual-channel miniature optical
transmitter that is fabricated with 850 nm VCSELs packaged in TOSA 
(Transmitter Optical Sub Assembly) format,
and the specialized LOCld65 laser driver \cite{IEEE2019} suitable for 
applications in high energy collider environments.
The LOCld65 is designed and fabricated with the TSMC 65 nm CMOS technology.
The first version of MTx+ assembly made with components specified for 10 Gbps
is evaluated for data transmission and configuration with an I$^2$C interface.
The speed performance is tested and reported for 14 Gbps ~\cite{MTXp_pixel2018}.

The revision of MTx+ aims for higher data transmission speed 
with the circuit board design and components upgraded to 25 Gbps.
In Sec.~\ref{sec:MTXp}, the design of MTx+ and the LOCld65 circuits are described.
The changes in revision and the test results at 25 Gbps are discussed 
in Sec.~\ref{sec:25G}. Comparison with modules assembled with 10~Gbps rated TOSAs 
is also conducted. A short summary   is 
discussed in Sec.~\ref{sec:Sum}.

\begin{figure}[ht!] 
  \vspace{.2cm}  
  \centering
  \includegraphics[width=1.00\linewidth]{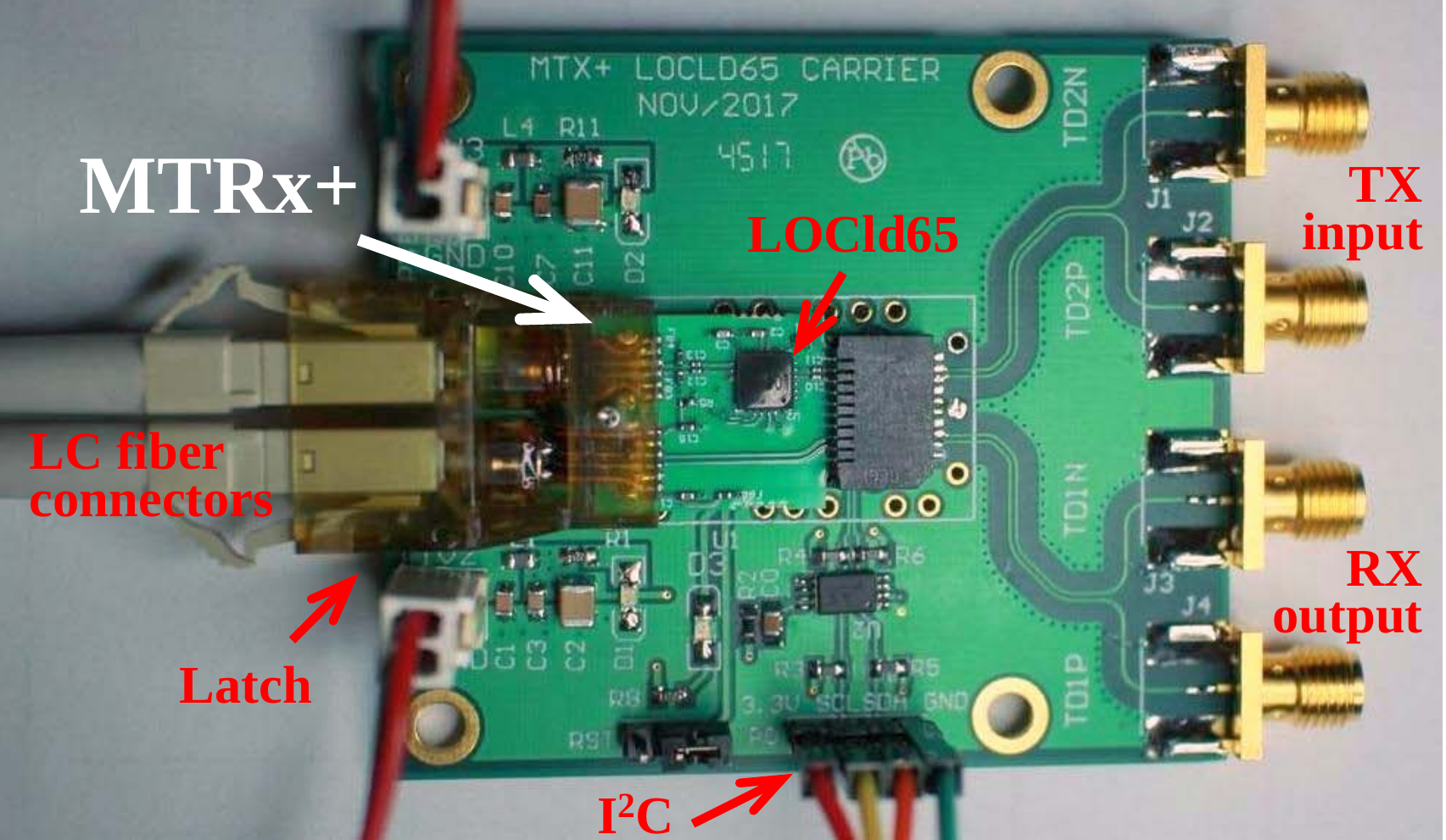}
  
  \vspace{+.2cm}
  \caption{ An earlier version of MTRx+ module with PCB made of FR-4 is shown.
  It is mounted on a test carrier board.
  The TOSA/ROSA are held in a customized latch for light coupling to LC 
  fiber connectors. The LC clips are rotated to the sides to reduce the 
  module height to 6 mm.
  The MTRx+ has the  PCB circuitry that mates to a SFP+ connector on the 
  carrier board.
  \label{fig:MTRxp_FR4} }
\end{figure}

\section{MTx+ transmitter with LOCld65}
\label{sec:MTXp}

The MTx+ is developed in continuation of the MTx transmitter 
for applications in the ATLAS Phase-I upgrade  \cite{MTx}.
The prototype MTx+ is designed with the circuit boards configured
for plug-in to SFP+ connector, and the electrical inputs in         
CML (Current Mode Logic) protocol for high-speed interfacing.
The differential signal swing is required for 100~mV minimum.
The module assembly is customized to a total height of 6 mm, 
with a latch holding TOSAs and LC type fiber connectors for light coupling.

Illustrated in Fig.~\ref{fig:MTRxp_FR4} is an earlier prototype module
mounted on a test carrier board.
The prototypes are also developed for the transceiver (MTRx+) 
with one transmitter (TX) channel assembled to the LOCld65.
The receiver (RX) channel 
is a customized ROSA (Receiver Optical Sub Assembly) 
that has the Photo-Detector current collected by 
the GBTIA receiver chip \cite{GBTIA} for output.  

The prototype modules were first prepared with components
specified for 10~Gbps including the TOSA, and the PCBs made of 
FR-4\footnote{
The FR-4 material is composed of fiber-glass 
reinforced epoxy-laminated sheets used in PCB manufacturing.
} material.
The speed performance of the transmitter is measured for 14~Gbps, 
which is limited by the laboratory test facility.  

\begin{figure}[t!]  
  \centering
  \includegraphics[width=1.00\linewidth]{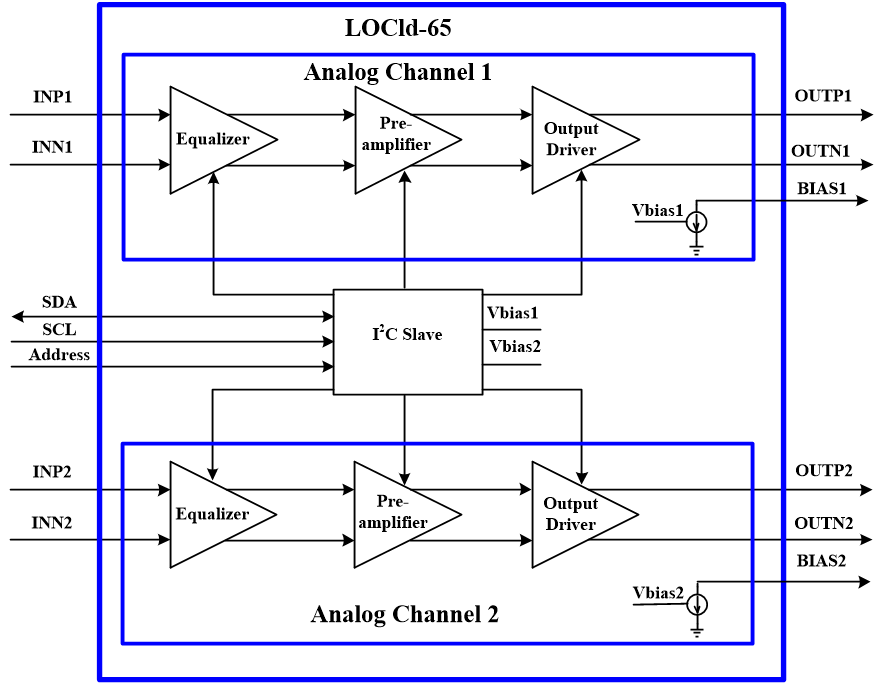}
  
  \caption{ Schematics of the dual-channel LOCld65 laser driver is illustrated.
  Each channel has a continuous-time equalizer, four stages of limiting 
  amplifiers, a high-current output driver, and a bias-current 
  generator. A common I$^2$C slave does the configuration to each
  functional block. 
  \label{fig:LOCld65} }
\end{figure}

The LOCld65 laser driver circuits developed for application
in radiation field is required for the minimum speed of 10~Gbps and is
configurable by an I$^2$C interface.
The schematic of the LOCld65 is plotted in Fig.~\ref{fig:LOCld65}.
It has two separate channels of the same circuits and a slave I$^2$C 
control section. 
The design goal is for each channel to amplify a differential signal of 
amplitude greater than 100~mV, and an 8~mA modulation current to a VCSEL.

Each of the LOCld65 channel consists of four functional blocks: an equalizer, 
four stages of limiting pre-amplifiers, 
a high-current output driver, and a bias-current generator. 
Each is programmable via the I$^2$C interface. 
The I$^2$C interface also provides an 1-bit register to turn the channel on or off 
independently.

The input of high-frequency signals is first compensated by the equalizer.
It is then amplified and driven into an 8~mA modulation current 
to the external VCSEL, which is biased at 6 mA by the bias-current generator.

The equalizer is implemented taking into account the attenuation of 
high-frequency components of the input signals induced by non-ideal 
factors in devices such as the PCB traces, bonding wires, and ESD diodes.
The equalizer design is a Continuous-Time Linear Equalizer with 
shunt peaking circuits to extend the bandwidth of input signals. 

The pre-amplifier is required  to output 
a large swing of up to 800 mV
 (peak to peak) to fully turn on/off the 
current driver which consists of an NMOS pair. 
A total gain of more than 18~dB is reached by designing the
output driver
with four stages of limiting amplifiers.
The bandwidth is extended with shared inductors 
and feedback that can be adjusted by the I$^2$C interface. 

The current driver is optimized for output connected by a flex cable
to a VCSEL in the
TOSA package. The modulation of the VCSEL current
is adjusted by a tail current source.

\begin{figure}[t!]    
  \vspace{.5cm}    
  \centering\includegraphics[width=1.\linewidth]{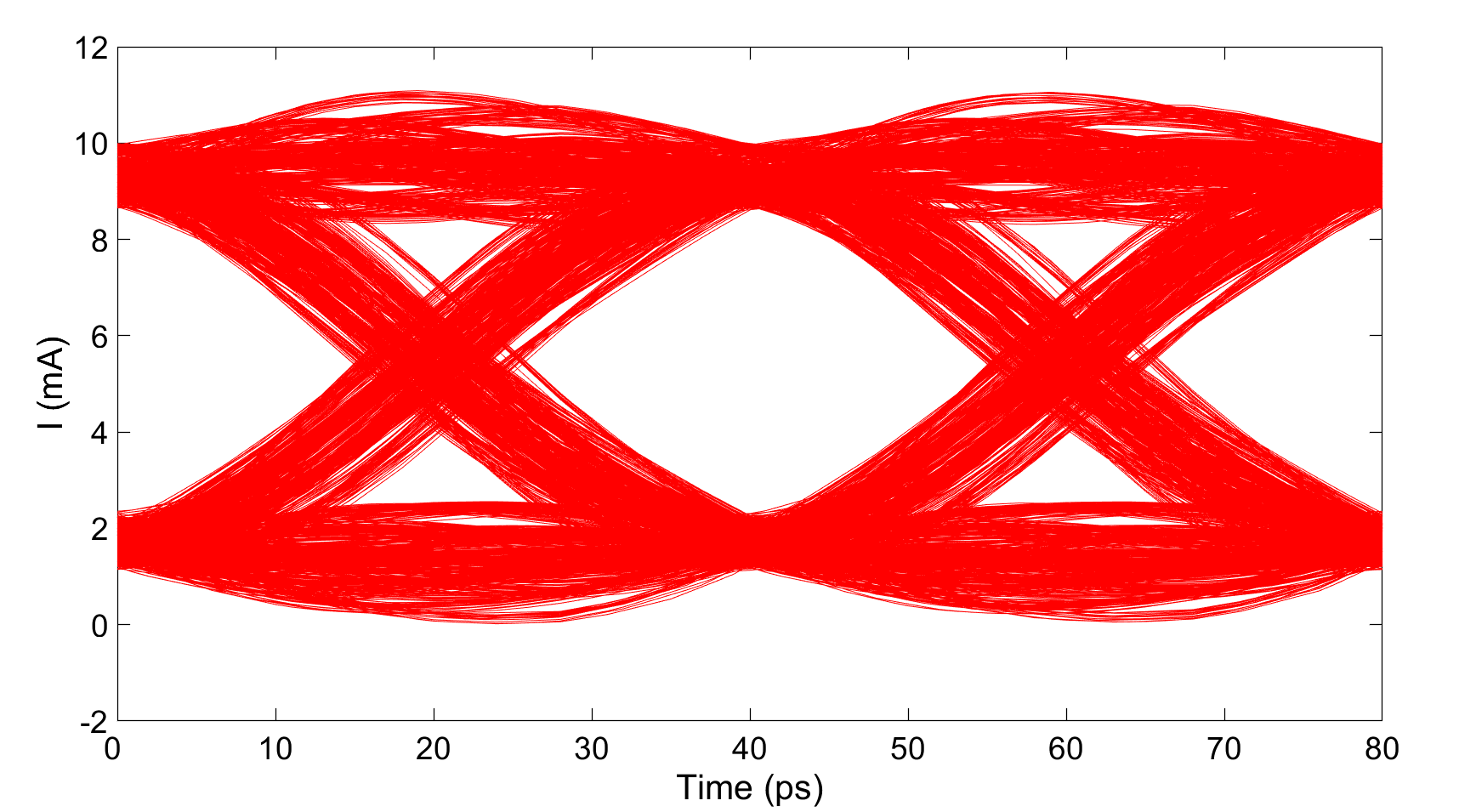}
  
  \vspace{+.0cm}
  \caption{ Post-layout simulation is conducted for the LOCld65 
  with differential input of 25 Gbps and the amplitude of 200~mV.
  The eye-diagram is plotted for the output current to VCSEL.
  \label{fig:LOCld_25G_simu} }
\end{figure}

The post-layout simulation is conducted for each stage of the LOCld65,
with the input amplitudes adjusted from the minimum required (100 mV)
to higher values.
The equalizer bandwidth can reach 32 GHz with a moderate R-C degeneration
setting. 
An eye-diagram simulation for the current driver output is illustrated
in Fig.~\ref{fig:LOCld_25G_simu}, for the input signal of 25 Gbps and
the amplitude of 200 mV. The eye-diagrams are compared for different 
input amplitudes. The results are compatible
for the amplitudes of 100 mV and up to 1~V.


\begin{figure}[b!]  
  \centering\includegraphics[width=1.\linewidth]{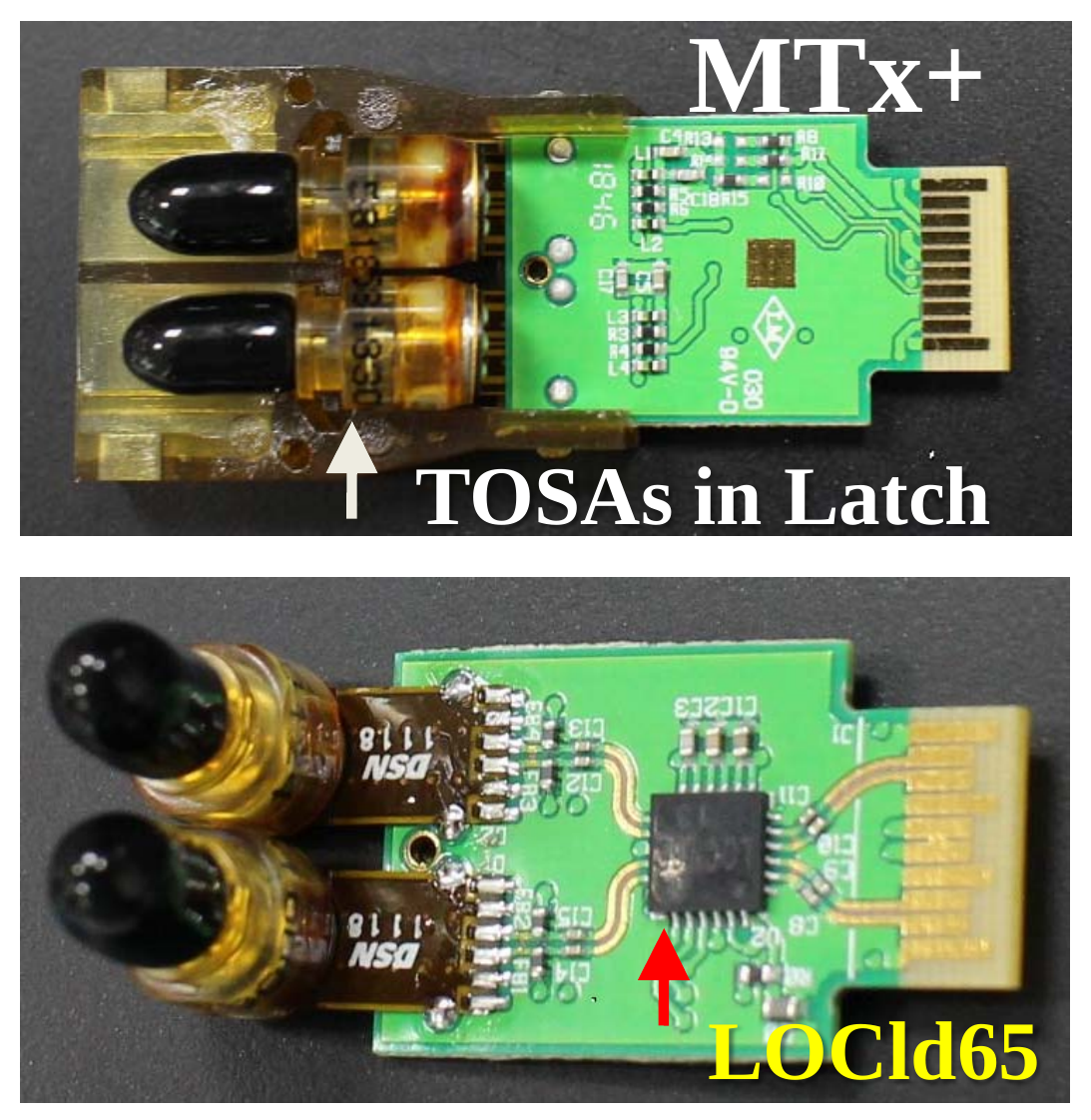}
  
  \vspace{+.0cm}
  \caption{ The assembly of a revised MTx+ transmitter is shown.
  It is made with the PCB of MEGTRON6 and the circuitry for plug-in to 
  SFP+ connector. The 25 Gbps TOSAs are clamped in a customized latch
  for light coupling to LC fiber connectors.
  \label{fig:MTxp_M6} }
\end{figure}


\begin{figure}[t!]  
  \vspace{.3cm}
  \centering\includegraphics[width=1.\linewidth]{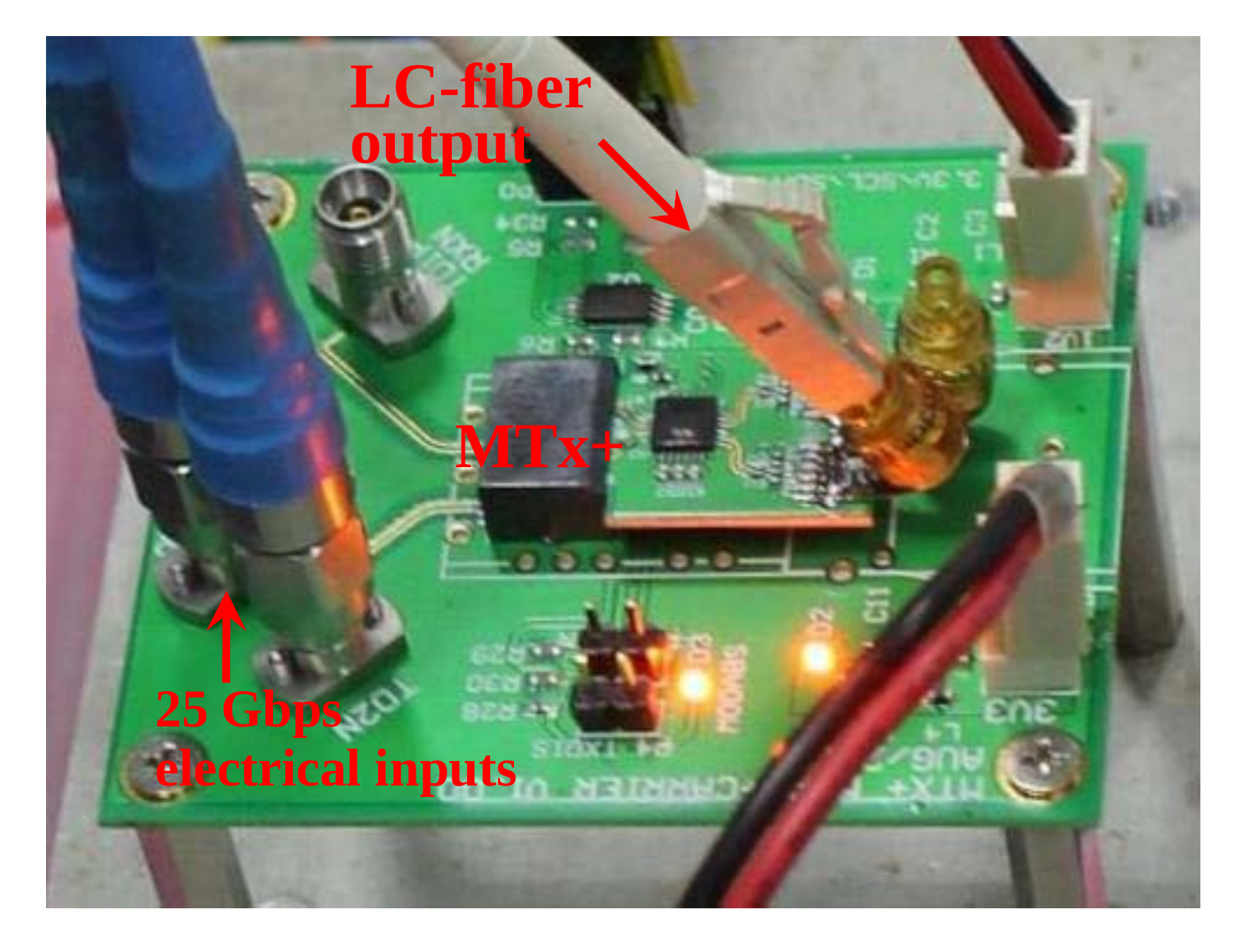}

  \caption{ The test setup is shown for a MTx+ mounted on a carrier board.
  The carrier board and the cabling for electrical input signals
  are made for 25 Gbps specification.
  \label{fig:25Gtest} }
\end{figure}

\begin{figure}[b!]
  \centering\includegraphics[width=1.\linewidth]{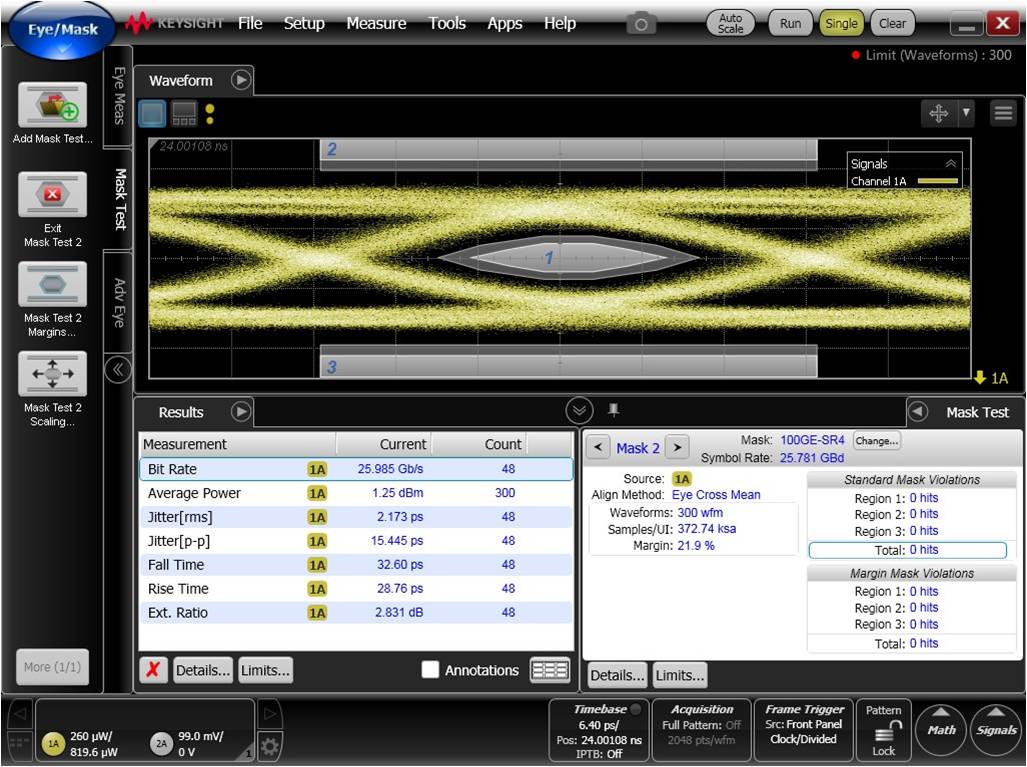}

  \caption{ The optical eye-diagram at 25 Gbps is shown for a MTx+ transmitter 
  channel assembled with a SAN-U TOSA specified for 25~Gbps.
  The gray area shows the 25 Gbps mask and the margin obtained is 22~\%. 
  \label{fig:eye25G} }
\end{figure}

\section{Revision of the MTx+ for 25 Gbps } 
\label{sec:25G}

The bandwidth of the initial version of MTx+ has exceeded the expected
10 Gbps speed.
It is investigated further for higher bandwidth
with the circuit design and components revised for 25~Gbps.
The PCB material is replaced by the Panasonic MEGTRON6 \cite{M6}. 
The circuit layout is modified and the cabling 
for electrical input signals are all upgraded accordingly.

The module assembly has used
two types of VCSELs in TOSA packages,
the 10 Gbps rated TOSA of Truelight \cite{TL} and 
the 25~Gbps TOSA of SAN-U \cite{SAN-U}.
Illustrated in Fig.~\ref{fig:MTxp_M6} are pictures of the new 
MTx+ modules. 
The test setup with the MTx+ mounted on a test carrier board is shown in 
Fig.~\ref{fig:25Gtest}. 
The electrical input waveforms are generated by an Anritsu MP1800A 
signal analyzer with the amplitude set to 1~V. 
The eye-diagrams are measured by a Keysight DCA-X 86100D oscilloscope.
For this test, input signals are tuned from 10~Gbps to 25~Gbps
in steps.
We optimize the operation of the LOCld65
equalizer receiving the input signals, as well as the bias and modulation to
the VCSELs, through the I$^2$C interface.
The same configuration is applied for all measurements at different speeds.

Shown in Fig.~\ref{fig:eye25G} is a typical eye-diagram measured at 25~Gbps
for a MTx+ channel assembled with a SAN-U TOSA.
This type of TOSA has a large light power. The measured value
is 1.25~dBm. 
The rise and fall time of the eye-diagram are about 30~ps 
at 10~\% and 90~\% of 
the step height, respectively, and the RMS jitter is 2~ps.
The gray area in the eye-diagram is the 25~Gbps mask. 
The mask margin obtained is 22~\%. 

\begin{figure}[h!]  
  \centering\includegraphics[width=1.\linewidth]{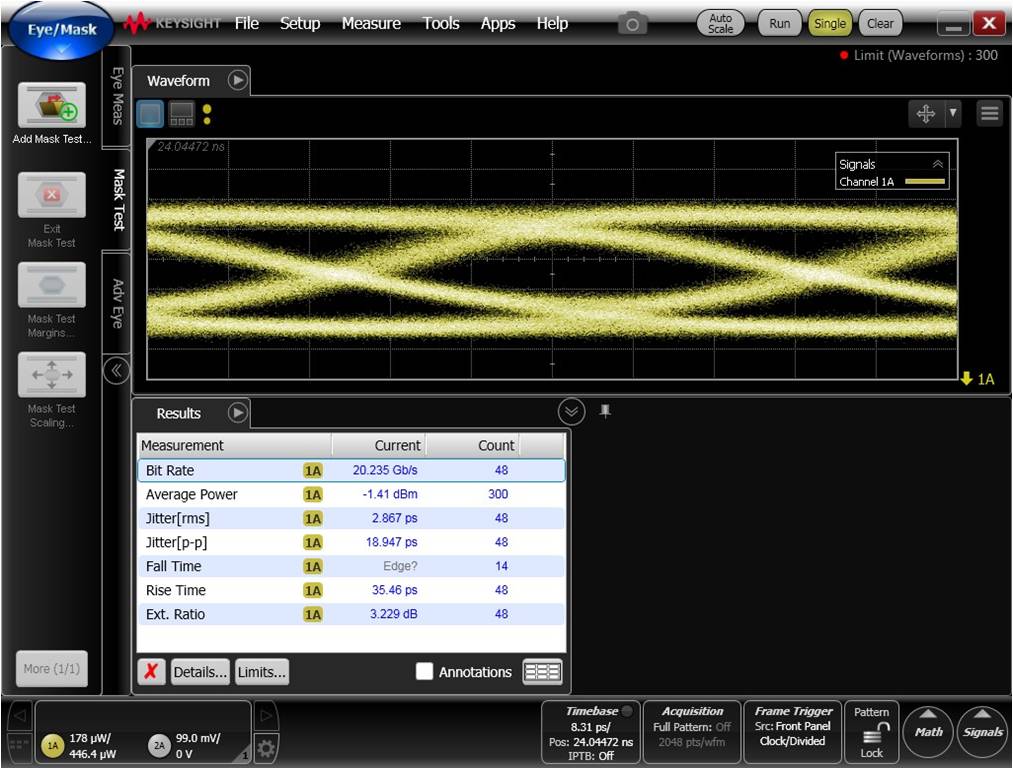}

  \caption{ The optical eye-diagram at 20 Gbps is shown for  
  an MTx+  
  transmitter channel assembled with a Truelight 10 Gbps TOSA.
  The test at higher speed has the eye-diagram blurred 
  and failed for data transmission.
  \label{fig:eye20G} }
\end{figure}

We have compared this previous measurement to the MTx+
using a slower 10 Gbps TOSA (of Truelight).
The bandwidth is limited by the TOSA and the test
has achieved a data transmission rate of 20~Gbps.
The eye-diagram is shown in Fig.~\ref{fig:eye20G}.
This confirms that the LOCld65 and the PCB assembly are 
all qualified for speed performance higher than 20~Gbps.

\section{Summary} 
\label{sec:Sum}
 
The dual-channel MTx+ transmitter consists of the LOCld65 laser driver
and 850 nm VCSELs in  TOSA packages. 
The speed performance is investigated with the modules made of 
PCB material and components qualified for 25 Gbps applications.
The eye-diagram observed for data transmission
at 25 Gbps has  a mask  margin measured to be 22~\%.
The revised MTx+ has reached the speed performance
of 25 Gbps being pursued.



{}

\end{document}